\documentclass[aps,prd,reprint,superscriptaddress,amsmath,amssymb]{revtex4-2}

\usepackage{graphicx}
\usepackage{dcolumn}
\usepackage{bm}

\newcommand{\mDP}{m_\text{DP}}
\newcommand{\PDP}{P_\text{DP}}
\newcommand{\Aeff}{A_\text{eff}}
\newcommand{\Dnu}{\Delta\nu}
\newcommand{\vE}{v_\text{E}}

\newcommand{\ueV}{\mu\text{eV}}
\newcommand{\GeV}{\text{GeV}}

\newcommand{\kHz}{\text{kHz}}
\newcommand{\MHz}{\text{MHz}}
\newcommand{\GHz}{\text{GHz}}
\newcommand{\mm}{\text{mm}}

\newcommand{\sqcm}{\text{cm}^2}
\newcommand{\cubcm}{\text{cm}^3}
\newcommand{\K}{\text{K}}
\newcommand{\kmpers}{\text{km}/\text{s}}

\newcommand{\range}[2]{\text{#1--}\allowbreak\text{#2}}

\newcommand{\eq}[1]{Eq.~(\ref{eq:#1})}

\newcommand{\fig}[1]{Fig.~\ref{fig:#1}}
\newcommand{\Fig}[1]{Figure~\ref{fig:#1}}
\newcommand{\tab}[1]{Table~\ref{tab:#1}}
\newcommand{\App}[1]{Appendix~\ref{app:#1}}

\begin{document}


\title{Search for Dark Photon Dark Matter at $\range{74}{159}\,\ueV/c^2$ Using Broadband Cryogenic Receivers}



\author{Hiroki~Takeuchi}
\email{takeuchi.hiroki.74w@st.kyoto-u.ac.jp}
\affiliation{Department of Physics, Faculty of Science, Kyoto University, Kyoto 606-8502, Japan}

\author{Shunsuke~Adachi}
\altaffiliation[Also at ]{High Energy Accelerator Research Organization (KEK), Tsukuba, Ibaraki 305-0801, Japan}
\affiliation{Department of Physics, Okayama University, Okayama 700-8530, Japan}

\author{Shunki~Endo}
\affiliation{Division of Physics, Faculty of Pure and Applied Sciences, University of Tsukuba, Tsukuba, Ibaraki 305-8571, Japan}

\author{Keigo~Hanimura}
\affiliation{Department of Physics, Faculty of Science, Kyoto University, Kyoto 606-8502, Japan}

\author{Michinori~Hirose}
\affiliation{Department of Mechanical and Electrical Engineering, Faculty of Engineering, Suwa University of Science, Chino, Nagano 391-0292, Japan}

\author{Shunsuke~Honda}
\altaffiliation[Also at ]{Tomonaga Center for the History of the Universe (TCHoU), University of Tsukuba, Tsukuba, Ibaraki 305-8571, Japan}
\affiliation{Division of Physics, Faculty of Pure and Applied Sciences, University of Tsukuba, Tsukuba, Ibaraki 305-8571, Japan}

\author{Yuki~Sakurai}
\affiliation{Department of Mechanical and Electrical Engineering, Faculty of Engineering, Suwa University of Science, Chino, Nagano 391-0292, Japan}

\author{Junya~Suzuki}
\affiliation{Department of Physics, Faculty of Science, Kyoto University, Kyoto 606-8502, Japan}

\author{Osamu~Tajima}
\altaffiliation[Also at ]{High Energy Accelerator Research Organization (KEK), Tsukuba, Ibaraki 305-0801, Japan}
\altaffiliation[Also at ]{Kavli Institute for the Physics and Mathematics of the Universe (WPI), UTIAS, The University of Tokyo, Kashiwa, Chiba 277-8583, Japan}
\affiliation{Department of Physics, Faculty of Science, Kyoto University, Kyoto 606-8502, Japan}

\collaboration{DOSUE-RR Collaboration}
\noaffiliation

\date{\today}

\begin{abstract}
Dark photons are hypothetical massive vector bosons that couple to ordinary photons through kinetic mixing $\chi$.
Dark photon dark matter (DPDM) converts into photons at a metal surface, with a frequency set by its mass.
Using cryogenic K- and Ka-band receivers with a broadband spectrometer, we searched for this signal over $74$--$159\,\ueV/c^2$ ($18$--$38.5\,\GHz$) with an integration time of $4.8\times10^5\,\mathrm{s}$ per frequency bin.
We found no statistically significant signal and set upper limits of $\chi<(\range{3.1}{17})\times10^{-12}$ at the 95\% confidence level over 99.8\% of the search range, assuming DPDM constitutes all local dark matter.
These results improve on previous limits by more than an order of magnitude over most of $\range{74}{110}\,\ueV/c^2$ and provide the first direct limits over most of $\range{110}{159}\,\ueV/c^2$, approximately two orders of magnitude below cosmological bounds.
\end{abstract}


\maketitle


Astrophysical and cosmological observations provide evidence for dark matter through its gravitational effects, but its microscopic nature remains unknown.
The dark photon, a hypothetical massive vector boson that mixes kinetically with the photon~\cite{Holdom1986,Caputo2021}, is a candidate for dark matter.
Dark photon dark matter (DPDM) produced by quantum fluctuations during inflation can account for the full dark matter abundance~\cite{Graham2016}.
For inflationary scales near the upper limit from cosmic microwave background observations~\cite{BICEPKeck2021}, this requires masses of order $100\,\ueV/c^2$.

The kinetic mixing of DPDM is constrained over a broad mass range by cosmological observations~\cite{Arias2012,Caputo2021}.
Direct searches set stronger limits~\cite{Tomita2020,Kotaka2023,Adachi2024,Bajjali2023,Knirck2024,Egge2025},
but only over narrow mass ranges, leaving most masses above $110\,\ueV/c^2$ without direct limits.
Resonant enhancement improves sensitivity but narrows the bandwidth of each configuration, so that broad coverage requires many tuning steps~\cite{Egge2025,Cervantes2022}.

We use a broadband detection method based on the conversion of DPDM into ordinary photons at a metal surface through kinetic mixing $\chi$~\cite{Horns2013}.
The photons are emitted nearly perpendicular to the surface~\cite{JaeckelKnirck2016}, which in our setup is a metal plate facing a horn antenna.
The photon frequency $\nu$ satisfies $h\nu\simeq\mDP c^2$, where $\mDP$ is the dark photon mass, $h$ is Planck's constant, and $c$ is the speed of light.
The signal has a fractional linewidth $\Dnu/\nu\sim10^{-6}$ due to the velocity dispersion of a virialized Galactic halo~\cite{Tomita2020}.

The total signal power $\PDP$ received by the antenna is given by~\cite{Horns2013}
\begin{equation}
    \PDP=\chi^2 c\rho\Aeff\alpha^2.
    \label{eq:chi}
\end{equation}
We assume DPDM constitutes all local dark matter, with energy density $\rho=(0.39\pm0.03)\,\GeV/\cubcm$~\cite{CatenaUllio2010}.
For randomly oriented DPDM polarization, we take the polarization factor $\alpha=\sqrt{1/3}$ for a single-polarization receiver~\cite{Caputo2021}.
The effective detection area $\Aeff(\nu)$ is the ratio of received signal power to power emitted per unit area by a perfectly conducting plane in the detected polarization.

In this Letter, we search for DPDM over $\range{74}{159}\,\ueV/c^2$ ($\range{18}{38.5}\,\GHz$) by extending the cryogenic receiver system of our previous search~\cite{Kotaka2023} to higher frequencies and employing dSpec,
a broadband spectrometer developed in Ref.~\cite{Takeuchi2025}.
Compared with the spectrum analyzer used in our previous searches~\cite{Kotaka2023,Adachi2024}, dSpec increases the integration time per frequency bin by a factor of 2000 for the same observing time and frequency coverage.
We improve on the previous limits by more than an order of magnitude over most of $\range{74}{110}\,\ueV/c^2$ and obtain the first direct limits across most of $\range{110}{159}\,\ueV/c^2$.

\begin{figure}[htb]
\includegraphics[width=8.6cm]{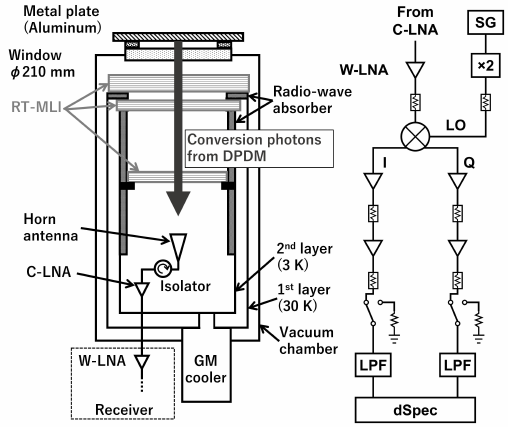}
\caption{\label{fig:setup}
Schematic cross section of the cryostat (left) and the warm receiver chain in the dashed box (right).
The horn antenna collects photons from the conversion of dark photon dark matter (DPDM) at the metal plate.
C-LNA/W-LNA: cold/warm low-noise amplifiers; RT-MLI: radio-transparent multilayer insulation~\cite{Choi2013};
SG: signal generator (FSL-0020); $\times2$: frequency doubler (SFA-203403205-KFSF-S1); LO: local oscillator; I/Q: in-phase/quadrature components; LPF: low-pass filter.
Switches select between signal paths and $50$-$\Omega$ terminations.
}
\end{figure}

\Fig{setup} shows the apparatus.
Alongside the K-band ($\range{18}{26}\,\GHz$) receiver used previously~\cite{Kotaka2023},
we employ a new Ka-band ($\range{26}{38.5}\,\GHz$) receiver (\tab{receiver}).
The Ka-band system noise temperature is typically $\range{60}{120}\,\K$, comparable to $\range{40}{100}\,\K$ in the K band.
The measurements were performed at ($\mathrm{35.03}^{\circ}$ N, $\mathrm{135.78}^{\circ}$ E),
with both horns pointing to the zenith and their polarization axes at an azimuth of approximately $165^\circ$,
measured clockwise from true north.

\begin{table}[htb]
\caption{\label{tab:receiver}Receiver components and frequency ranges.
IF: intermediate frequency.}
\begin{ruledtabular}
\begin{tabular}{lccc}
Receiver & K band & \multicolumn{2}{c}{Ka band} \\
Range ($\GHz$) & 18--26 & 26--32 & 32--38.5 \\
\colrule
Horn antenna & SGH-42-SC000 & \multicolumn{2}{c}{SAR-2507-28-S2} \\
C-LNA & \multicolumn{3}{c}{LNF-LNC15-29B} \\
W-LNA & ALM-1826S210 & \multicolumn{2}{c}{ZVA-24443G1+} \\
Mixer & \multicolumn{3}{c}{MMIQ-1040LS} \\
IF amplifier & ZX60-43-S+ & ZX60-43-S+ & ZKL-2R7+ \\
\end{tabular}
\end{ruledtabular}
\end{table}

The downconverted intermediate-frequency (IF) signal is recorded by dSpec, which continuously acquires spectra over $4\,\GHz$ with $15.6$-kHz bin spacing and a $0.983$-s integration time per spectrum.
The observed band, centered on the local oscillator (LO) frequency, is selected by tuning the LO.

We evaluated $\Aeff(\nu)$ by numerically propagating the conversion fields through the apparatus geometry and calculating their coupling to each horn, whose aperture fields were obtained from electromagnetic simulations.
Conversion photons from other metal parts inside the apparatus interfere with those from the plate with uncertain relative phases.
We conservatively adopted the minimum $\Aeff$ over these phases and over the uncertain geometry below the window, which ranges over $\range{31.6}{63.7}\,\sqcm$ and $\range{24.5}{52.7}\,\sqcm$ for the K- and Ka-band receivers, respectively.
The calculation, boundary conditions, and systematic uncertainties are described in \App{aeff}.

The search covered $\range{18}{38.5}\,\GHz$ using the LO frequency pairs listed in \tab{measurements}.
For each pair, we alternated $600$-s measurements between the two LO frequencies, separated by $20\,\MHz$, to investigate spurious signals and to retain usable data where only one of them was contaminated.
Between these measurements, we switched the spectrometer inputs to $50$-$\Omega$ terminations for $10\,\mathrm{s}$ to probe spurious signals originating downstream of the switches.
The total integration time for the search was $2.4\times10^5\,\mathrm{s}$ at each LO frequency.

\begin{table}[htb]
\caption{\label{tab:measurements}Search frequency ranges, LO frequency pairs, and measurement periods.}
\begin{ruledtabular}
\begin{tabular}{lccc}
Receiver & K band & \multicolumn{2}{c}{Ka band} \\
Range ($\GHz$) & 18--26 & 26--32 & 32--38.5 \\
\colrule
LO pairs & $(20.00,20.02)$ & $(27.50,27.52)$ & $(33.50,33.52)$ \\
($\GHz$) & $(22.25,22.27)$ & $(30.50,30.52)$ & $(36.50,36.52)$ \\
 & $(24.50,24.52)$ & & \\
Period & Sep.~27--Oct.~22 & Dec.~18--30 & Jan.~16--28 \\
 & 2025 & 2025 & 2026 \\
\end{tabular}
\end{ruledtabular}
\end{table}

At the lower frequency of each LO pair, we also performed a separate measurement with the metal plate replaced by a room-temperature absorber, with an integration time exceeding $2.4\times10^5\,\mathrm{s}$.
These data were used as mock search data and for characterizing spurious signals.

To calibrate the search data, we measured the receiver gain using the hot--cold method~\cite{Partridge1995}.
Radio-wave absorbers at room and liquid-nitrogen temperatures were placed above the cryostat window, and spectra were recorded for $15\,\mathrm{s}$ at each temperature.
During the search, we performed these measurements daily at each LO frequency.
Separately, we repeated the hot--cold measurement hourly over 24 hours to assess gain stability.
In the analysis, we used the gains from the daily measurements, averaged in linear units at each frequency.

For each LO frequency, we averaged the spectra over the full measurement period and estimated the statistical fluctuations of this time-averaged spectrum using null spectra, as in our previous searches~\cite{Kotaka2023,Adachi2024}.
We generated 10,000 null spectra by pairing consecutive spectra, reversing the sign of one randomly chosen spectrum in each pair, and averaging all spectra.
This cancels stable spectral features while preserving the statistical noise level of the time-averaged spectrum.
At each frequency bin, the standard deviation across the null spectra was used as the statistical uncertainty for both the time-averaged spectrum and every null spectrum.

\begin{figure}[tb]
\centering
\includegraphics[width=8.6cm]{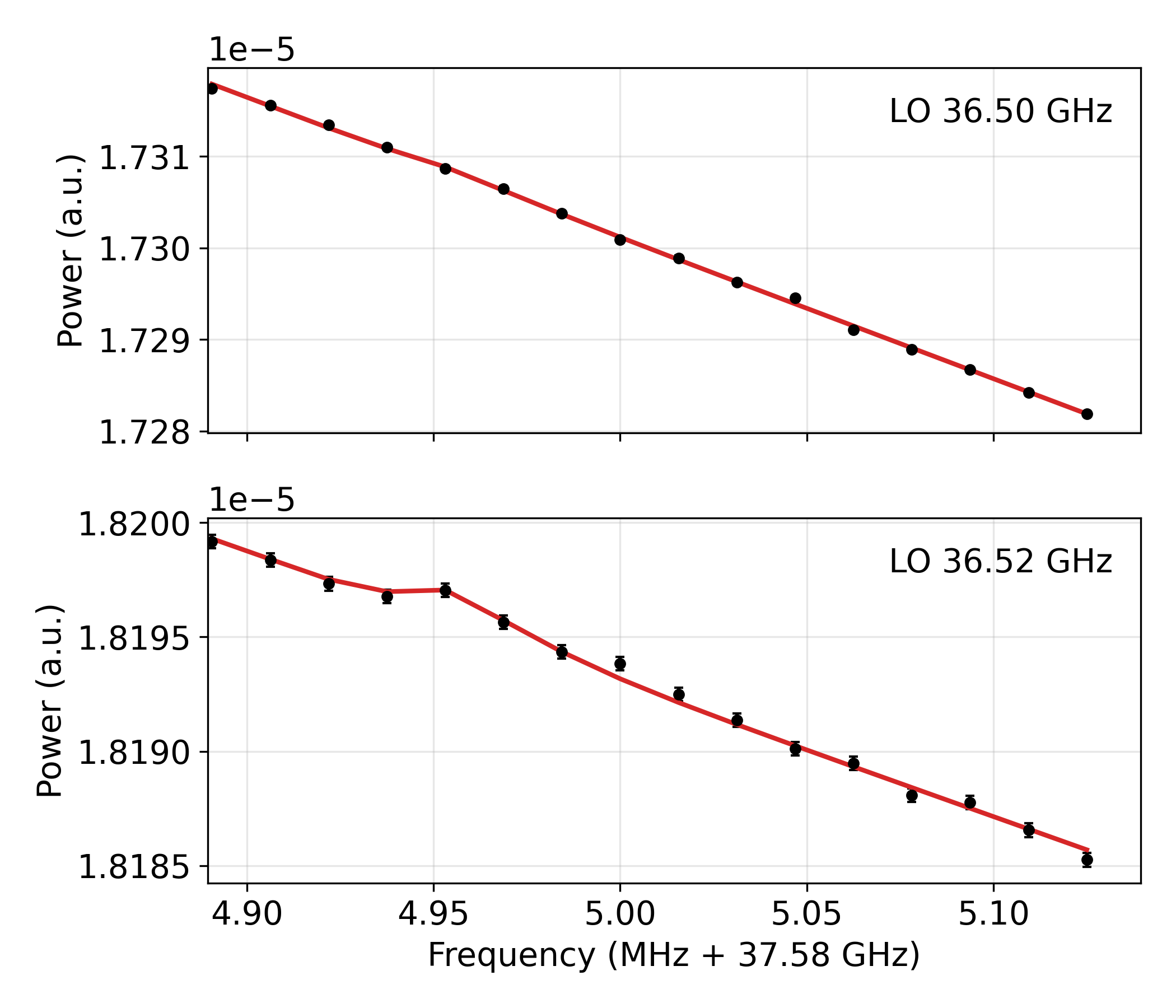}
\caption{
Measured spectra near $\nu_0=37.584939\,\GHz$ and fits for signal extraction (solid lines).
The upper and lower panels correspond to LO frequencies of $36.50$ and $36.52\,\GHz$, respectively.
Error bars indicate $1\sigma$ statistical uncertainties.
At this frequency, we obtained the smallest local $p$ value in the search, while the corresponding global $p$ value indicated no significant excess.
}
\label{fig:fit_spectra}
\end{figure}

As in our previous searches~\cite{Kotaka2023,Adachi2024}, we extracted the signal power for each trial dark photon mass, expressed as $\nu_0\equiv\mDP c^2/h$, by fitting the expected signal shape with a linear background:
\begin{equation}
f(\nu;P,a,b)=P f_{\rm sig}(\nu;\nu_0)+a(\nu-\nu_0)+b.
\label{eq:fit}
\end{equation}
Here, $\nu$ is the center frequency of each FFT bin, $P$ is the fitted signal power with statistical uncertainty $\Delta P$, and $a$ and $b$ describe the background.
The signal function $f_{\rm sig}$ gives the fraction of total signal power expected in each bin.
It was calculated from the Maxwell--Boltzmann halo spectrum with Galactic circular speed and Earth speed both set to $220\,\kmpers$~\cite{Drukier1986,Tomita2020}, convolved with the power response of the rectangular-window FFT in dSpec~\cite{Takeuchi2025}.
We evaluated $P$ at $\nu_0$ values spaced by $1.95\,\kHz$, using $\chi^2$ fits to the time-averaged and null spectra over $\nu_0-50$ to $\nu_0+200\,\kHz$.

We applied the gain calibration to $P$ and $\Delta P$.
We then combined the results from the two LO frequencies with inverse-variance weights to obtain the received signal power $\PDP$ and its statistical uncertainty $\Delta\PDP$.
Data affected by spurious signals were omitted from this combination.

We evaluated two classes of spurious signals that could mimic a DPDM signal: those originating in the mixer, LO, or downstream receiver chain (IF spurious signals), and those appearing at fixed radio frequencies (RF spurious signals).
The former were identified using absorber and termination measurements and comparisons of spectra at different LO frequencies.
The latter were traced to harmonics of the dSpec clocks in follow-up measurements with an independent spectrum analyzer: the tested peaks appeared only when dSpec was powered on.
Data affected by either class were excluded from the search.
The identification procedures and exclusion criteria are described in \App{analysis}.
The same selection, calibration, and combination were applied to the null spectra.

\begin{figure}[tb]
\centering
\includegraphics[width=8.6cm]{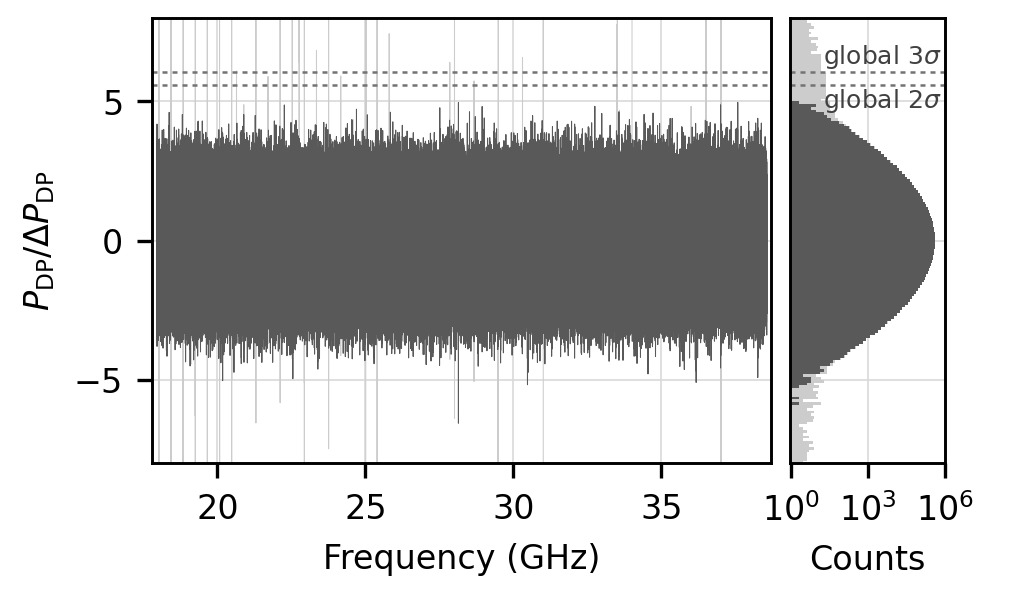}
\caption{
The ratio $\PDP/\Delta\PDP$ of extracted signal power to its statistical uncertainty is plotted versus frequency (left) and as a histogram (right).
Light gray indicates results from intervals excluded from the search.
Dashed lines indicate global significances of $2\sigma$ and $3\sigma$.
}
\label{fig:search_statistics}
\end{figure}

After the exclusion of frequency intervals contaminated by identified spurious signals, 99.8\% of the search range remained.
We evaluated the statistical significance following the procedure used in our previous searches~\cite{Kotaka2023,Adachi2024}.
The fits to the null spectra provided the distributions of the ratio $\PDP/\Delta\PDP$ under the zero-signal hypothesis.
These distributions were consistent with a standard normal distribution.
Using these distributions, we calculated the local $p$ value at each frequency as the probability of obtaining a ratio at least as large as the observed value.

We accounted for the look-elsewhere effect~\cite{Foster2018} using an effective number of independent trials, $N_{\rm eff}=1.94\times10^6$, estimated from the null spectra with correlations between neighboring frequencies included.
The probability of an equally or more significant fluctuation anywhere in the search range was
\begin{equation}
p_{\rm global}=1-(1-p_{\rm local}^{\rm min})^{N_{\rm eff}},
\label{eq:pglobal}
\end{equation}
where $p_{\rm local}^{\rm min}$ is the smallest local $p$ value.

\Fig{search_statistics} shows the frequency dependence and distribution of $\PDP/\Delta\PDP$ in the search data.
The smallest local $p$ value was $2.43\times10^{-7}$ at $\nu_0=37.584939\,\GHz$ ($\PDP/\Delta\PDP=4.96$), corresponding to a global $p$ value of $0.38$.
\Fig{fit_spectra} shows the spectra and fits for both LO frequencies at this $\nu_0$.
No statistically significant DPDM signal was found.

The systematic uncertainties in $\chi$ are summarized in \tab{syst}.
The uncertainties in $\Aeff$ were evaluated by recalculating $\Aeff$ with varied window geometry and photon direction.
We compared the frequency-averaged linear gain with its time average at each LO frequency, separately for the daily calibrations during the search and the 24-hour measurements.
We adopted the largest relative deviations, $9.5\%$ (K band) and $5.3\%$ (Ka band), as the gain uncertainties; these contribute $5.1\%$ and $2.8\%$, respectively, to the uncertainty in $\chi$.
The uncertainty in the absorber temperature contributes $0.5\%$.

We evaluated the bias in $P$ due to the discrete spacing of $\nu_0$ by fitting noiseless signals at frequencies between adjacent $\nu_0$ values and comparing the fitted and injected powers.
We evaluated the effect of Earth speed variation by varying $\vE$ in the signal model, following our previous analysis~\cite{Adachi2024}.
The uncertainty in the local dark matter density~\cite{CatenaUllio2010} contributes $3.9\%$ to that in $\chi$.

\begin{table}[htb]
\caption{\label{tab:syst}
Systematic uncertainties in $\chi$.
Interference with conversion photons from other metal parts and the geometry below the window are accounted for by adopting the minimum $\Aeff(\nu)$ and are not listed.
}
\begin{ruledtabular}
\begin{tabular}{lcc}
Receiver & K band & Ka band \\
 & (\%) & (\%) \\
\colrule
Effective detection area ($\Aeff$) & 1.6--23.6 & 2.2--24.9 \\
\quad Window geometry & 1.5--23.6 & 2.0--24.9 \\
\quad Direction of conversion photons & 0.5 & 0.8 \\
Gain & 5.1 & 2.8 \\
\quad Time variation & 5.1 & 2.8 \\
\quad Absorber temperature & \multicolumn{2}{c}{0.5} \\
$\nu_0$ spacing & \multicolumn{2}{c}{0.8} \\
Earth speed variation ($\vE$) & \multicolumn{2}{c}{0.9} \\
Dark matter density ($\rho$) & \multicolumn{2}{c}{3.9} \\
\colrule
Total & 6.8--24.5 & 5.4--25.4 \\
\end{tabular}
\end{ruledtabular}
\end{table}

\begin{figure}[tb]
\centering
\includegraphics[width=8.6cm]{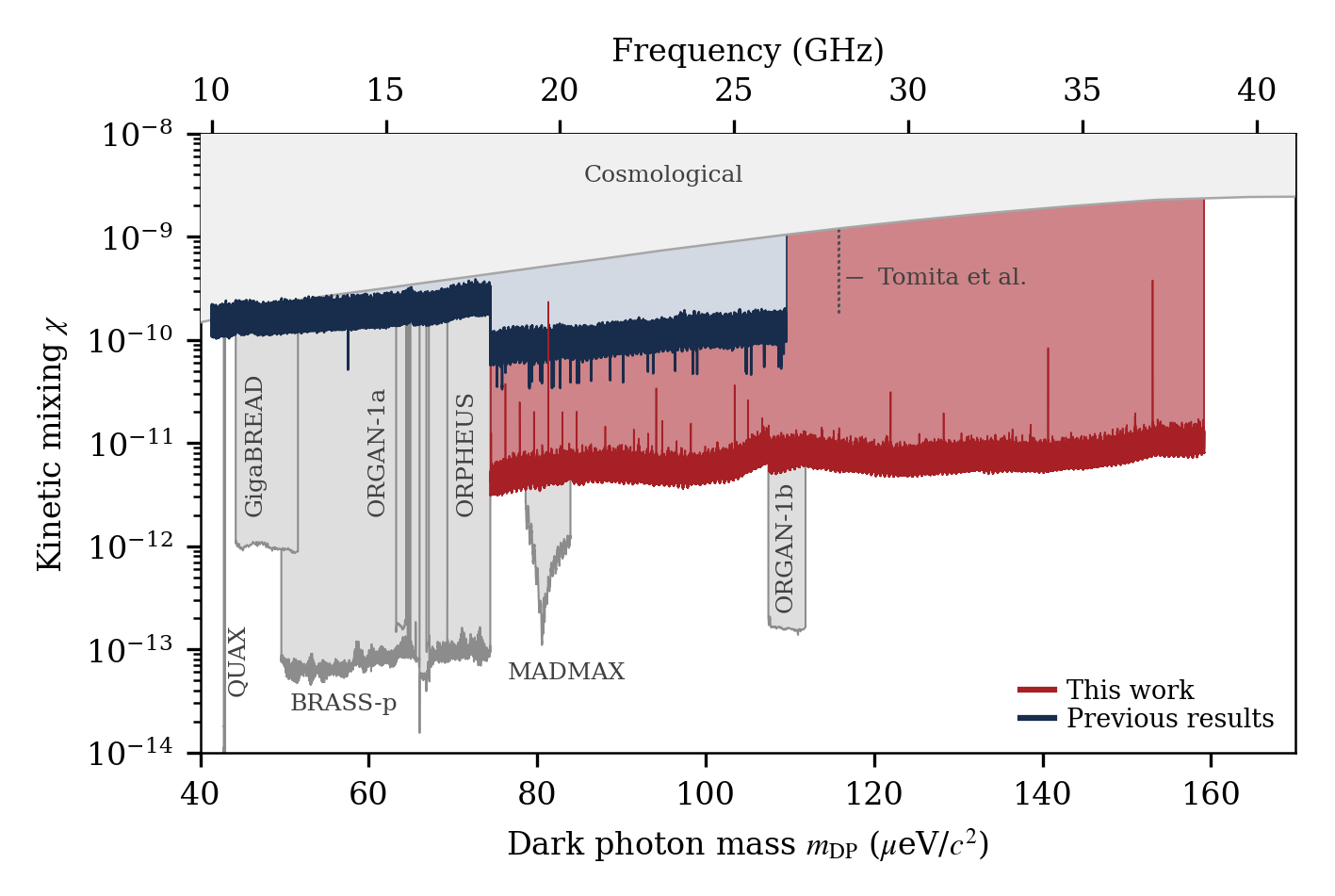}
\caption{\label{fig:limits}
Upper limits on $\chi$ at 95\% confidence level from this work (red) and our previous searches~\cite{Kotaka2023,Adachi2024} (blue).
The limits from this work include systematic uncertainties.
Shading above the curves indicates excluded regions.
Other experimental~\cite{Tomita2020,Knirck2024,Bajjali2023,Cervantes2022,Egge2025,Quiskamp2022,Quiskamp2024,Alesini2021,DiVora2023,Caputo2021}
and cosmological~\cite{Arias2012,Witte2020} constraints are shown in gray, using data from AxionLimits~\cite{AxionLimits}.
}
\end{figure}

We set the 95\% confidence-level upper limit on $\chi$ at each $\nu_0$ to the value for which the probability of $\PDP/\Delta\PDP\le\max[(\PDP/\Delta\PDP)_{\rm obs},0]$ is 5\%.
In this calculation, the expected signal power was given by \eq{chi} with the minimum $\Aeff(\nu)$, the statistical fluctuation was treated as Gaussian with standard deviation $\Delta\PDP$, and the total systematic uncertainty in \tab{syst} as an independent Gaussian factor in $\chi$.
\Fig{limits} shows the resulting 95\% confidence-level upper limits, $\chi<(\range{3.1}{17})\times10^{-12}$ over $\range{74}{159}\,\ueV/c^2$, together with previous constraints.
We also set limits in intervals excluded from the search, without subtracting identified spurious signals.
These conservative upper limits are as high as $3.8\times10^{-10}$.

We searched for DPDM over $\range{74}{159}\,\ueV/c^2$ using cryogenic K- and Ka-band receivers with a broadband spectrometer.
We found no statistically significant signal and set 95\% confidence-level upper limits of $\chi<(\range{3.1}{17})\times10^{-12}$.
These limits improve on our previous limits~\cite{Kotaka2023} by more than an order of magnitude over most of $\range{74}{110}\,\ueV/c^2$.
Over most of $\range{110}{159}\,\ueV/c^2$, they are the first direct limits, approximately two orders of magnitude below cosmological bounds.

\begin{acknowledgments}
This work was supported by JSPS KAKENHI Grants No.~JP20K14486, No.~JP20K20427, No.~JP21H05460, No.~JP22K18713, No.~JP24KJ1452, No.~JP24KK0066, and No.~JP25H02181;
JST CREST Grant No.~\mbox{JPMJCR23I3}, Japan;
JST FOREST Program Grant No.~\mbox{JPMJFR256K}, Japan;
and a research grant from the Murata Science and Education Foundation.
H.T. was supported by a JSPS Research Fellowship for Young Scientists.
S.A. acknowledges support from the Hakubi Project of Kyoto University.
\end{acknowledgments}

\appendix

\section{Effective detection area}
\label{app:aeff}

\begin{figure}[htb]
\centering
\includegraphics[width=8.6cm]{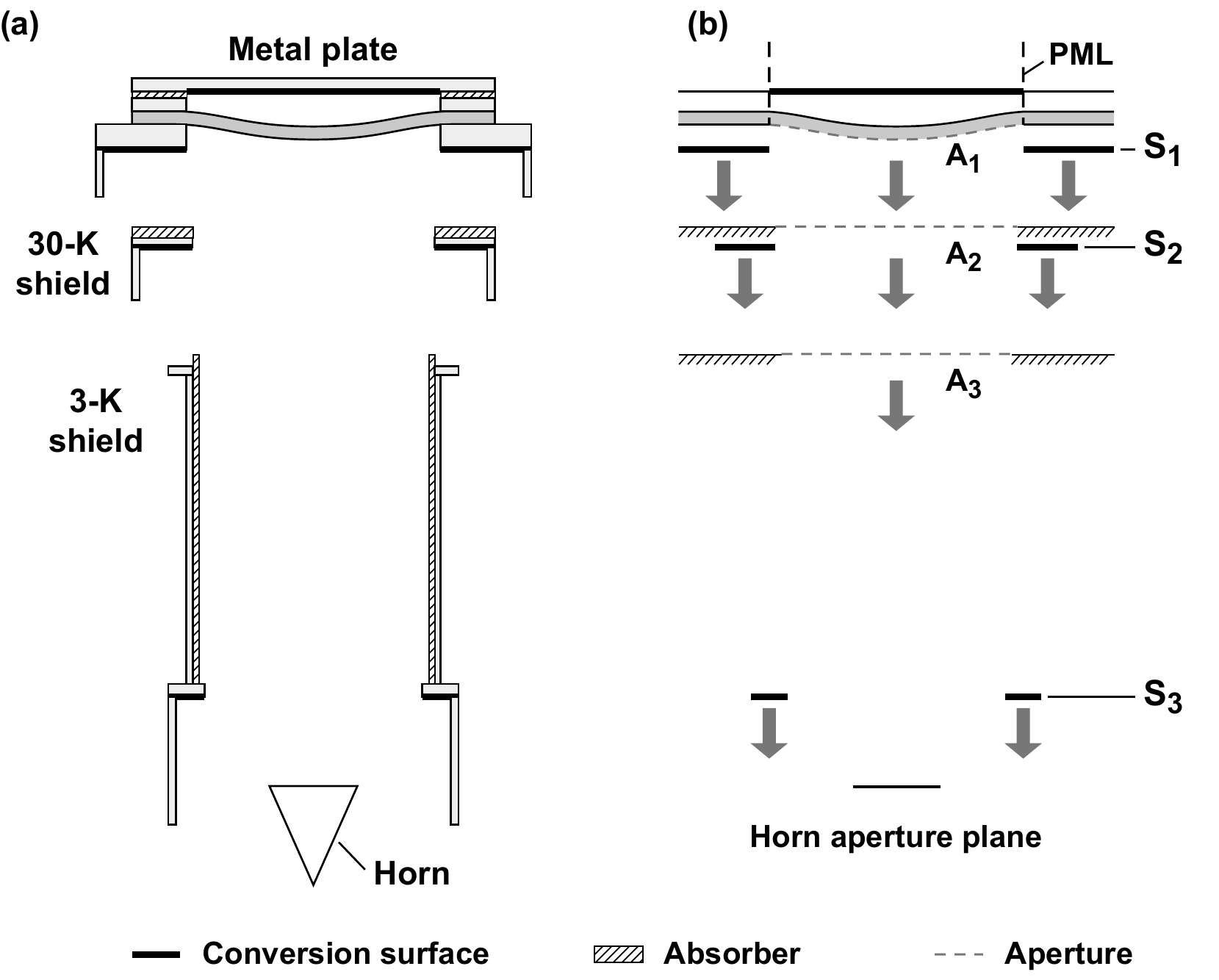}
\caption{
Geometry for the effective detection area calculation:
(a) schematic section of the cryostat and
(b) computational model, not to scale.
The gray layer is the vacuum window.
$S_1$--$S_3$ denote the exposed annular metal surfaces facing the horn on the vacuum-chamber ceiling,
the 30-K shield ceiling,
and the joint of the 3-K shield, respectively.
}
\label{fig:aeff_geometry}
\end{figure}

\Fig{aeff_geometry} shows the geometry used in the calculation.
The vacuum window of our previous search~\cite{Kotaka2023}, made of ultrahigh-molecular-weight polyethylene, was modeled with its measured profile, which is bowed inward by atmospheric pressure.
We calculated the conversion fields at the metal plate and window interfaces from the electromagnetic boundary conditions in the presence of DPDM~\cite{JaeckelRedondo2013}.
The plate was treated as a perfect conductor, and propagation between these surfaces, including multiple reflections,
was evaluated using equivalent electric and magnetic surface currents~\cite{Jackson1999}.
A perfectly matched layer (PML) provided a nonreflecting radial boundary outside the window opening.
Below the window, radio-wave absorbers suppress multiple reflections and limit the horn's view of surrounding structures.
From the window exit aperture $A_1$, we propagated the field through $A_2$ and $A_3$, modeled as openings in perfectly absorbing screens at the 30-K shield and the absorber-lined cylinder entrance, respectively.
Using electromagnetic reciprocity~\cite{Rumsey1954}, we calculated the received power from the conversion fields propagated to the horn aperture plane and the aperture fields of the horn in transmitting mode.
The latter were simulated with Ansys HFSS~\cite{HFSS} for unit incident power at the waveguide port.

The conversion fields, including those from the three exposed annular metal surfaces $S_1$--$S_3$ [\fig{aeff_geometry}(b)], were combined coherently because the apparatus lies within the spatial coherence length of Galactic DPDM~\cite{Caputo2021}.
Since the annuli belong to separate cryostat components at different temperatures, their positions relative to the plate cannot be determined with a precision much better than the photon wavelengths of $8$--$17\,\mm$.
We therefore obtained conservative bounds on $\Aeff$ by independently varying each annular contribution's phase over $0$--$2\pi$ relative to the plate--window field.
The minimum $\Aeff$ is lower than that without the annular contributions by up to 44.4\% and 41.5\% for the K- and Ka-band receivers, respectively.
Similarly, for the geometry below the window, namely the horn position and the dimensions of the apertures $A_2$ and $A_3$, we adopted the lowest $\Aeff$ within the uncertainty of each parameter.
We used the resulting minimum $\Aeff$ for the reported limits.
$\Aeff$ and its bounds were evaluated at 1-GHz intervals, at which the horn fields were simulated, and linearly interpolated in between.

The remaining uncertainties in $\Aeff$, from the window geometry and the direction of the conversion photons, were evaluated by recalculating $\Aeff$ with these varied; the resulting uncertainties in $\chi$ are summarized in \tab{syst}.

\section{Identification and exclusion of spurious signals}
\label{app:analysis}

In dSpec, a $122.88$-MHz crystal oscillator generates the $409.6$-MHz ADC reference clock.
IF spurious signals were identified from the absorber measurements by examining features with $|P/\Delta P|\ge5.5$ and comparing them with $50$-$\Omega$ termination spectra and measurements at different LO frequencies.
The dominant components occurred at integer multiples of $40.96\,\MHz$, as previously observed~\cite{Takeuchi2025}, and included harmonics of $409.6$, $122.88$, $655.36$, $1146.88$, and $1802.24\,\MHz$.
For each LO pair, we also excluded $40.96$-MHz multiples identified in the corresponding absorber data.

RF peaks at the known clock harmonics with $P/\Delta P\ge5.5$ were examined in follow-up measurements using an Anritsu MS2840A spectrum analyzer in place of dSpec, comparing spectra taken with dSpec powered on and off at a $300$-Hz resolution bandwidth over a $2.5$-MHz span.
The search data contained 19 such peaks at harmonics of $409.6\,\MHz$ and one at the 168th harmonic of $122.88\,\MHz$ ($20.64384\,\GHz$).
The tested $409.6$-MHz harmonics, $n=44$--49 in the K band and $n=72$ ($29.4912\,\GHz$) in the Ka band, and the $122.88$-MHz peak appeared only with dSpec powered on and were thus identified as spurious.
For the $409.6$-MHz series, which showed this behavior at consecutive harmonics, we excluded all harmonics in the search range, including the other 12 peaks and those below the threshold.
The $122.88$-MHz peak was excluded individually.

Intervals within $10\,\MHz$ of each LO frequency were excluded because of the low IF gain near DC.
Using the calculated response to monochromatic signals, we excluded intervals where $P$ would change by at least $\Delta P$ (\tab{spur_exclusion}).
Where data at only one LO frequency were affected, those at the other were retained; intervals excluded at both were omitted from the search.
The remaining intervals cover 99.8\% of the search range.

\begin{table}[tb]
\caption{\label{tab:spur_exclusion}
Frequency intervals excluded around identified spurious signals.
}
\begin{ruledtabular}
\begin{tabular}{cc}
$|P/\Delta P|$ & Frequency offset ($\MHz$) \\
\colrule
$[0,10)$        & $[-0.1,+0.3)$ \\
$[10,10^3)$     & $[-0.5,+0.5)$ \\
$[10^3,10^4)$   & $[-1,+1)$ \\
$\geq 10^4$ & $[-10,+10)$ \\
\end{tabular}
\end{ruledtabular}
\end{table}

\bibliography{references}

\end{document}